\documentclass[%
 reprint,
 superscriptaddress,
 amsmath,
 amssymb,
 aps,
 pra
]{revtex4-1}

\usepackage{graphicx}
\usepackage{dcolumn}
\usepackage{bm}
\usepackage{multirow}

\begin{document}

\title{Emergence of Magnetism in Bulk Amorphous Palladium}

\author{Isa\'{\i}as Rodr\'{\i}guez}
\affiliation{Facultad de Ciencias, Universidad Nacional Aut\'onoma de M\'exico, Apartado Postal 70-542, Ciudad Universitaria, CDMX, 04510, M\'exico}
\author{Renela M. Valladares}
\affiliation{Facultad de Ciencias, Universidad Nacional Aut\'onoma de M\'exico, Apartado Postal 70-542, Ciudad Universitaria, CDMX, 04510, M\'exico}
\author{David Hinojosa-Romero}
\affiliation{Instituto de Investigaciones en Materiales, Universidad Nacional Aut\'onoma de M\'exico, Apartado Postal 70-360, Ciudad Universitaria, CDMX, 04510, M\'exico}
\author{Alexander Valladares}
\affiliation{Facultad de Ciencias, Universidad Nacional Aut\'onoma de M\'exico, Apartado Postal 70-542, Ciudad Universitaria, CDMX, 04510, M\'exico}
\author{Ariel A. Valladares}
\email[Corresponding author; e-mail: ]{valladar@unam.mx}
\affiliation{Instituto de Investigaciones en Materiales, Universidad Nacional Aut\'onoma de M\'exico, Apartado Postal 70-360, Ciudad Universitaria, CDMX, 04510, M\'exico}

\date{\today}

\begin{abstract}
Magnetism in palladium has been the subject of much work and speculation. Bulk crystalline palladium is paramagnetic with a high magnetic susceptibility. Palladium under pressure and palladium nanoclusters have generated interest to scrutinize its magnetic properties. Here we report another possibility: Palladium may become an itinerant ferromagnet in the amorphous bulk phase at atmospheric pressure. Atomic palladium is a d$^{10}$ element, whereas bulk crystalline Pd is a d$^{10-x}$(sp)$^{x}$ material; this, together with the possible presence of ‘unsaturated bonds’ in amorphous materials, may explain the remnant magnetism reported herein. This work presents and discusses magnetic effects in bulk amorphous palladium.
\begin{figure}[h]
\includegraphics[width=0.7\textwidth]{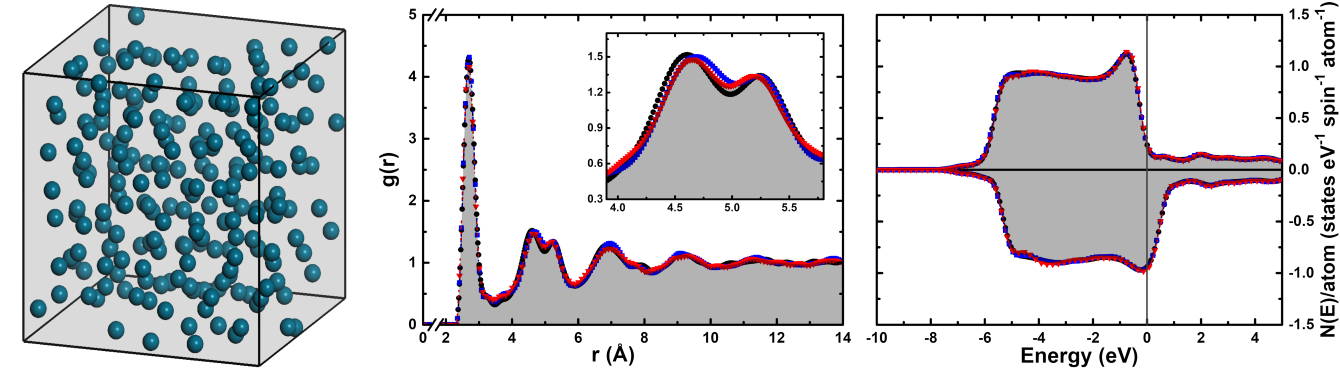}
\end{figure}
\end{abstract}

\keywords{Palladium, computer simulations, magnetism and disorder, ab initio techniques}

\maketitle

\section{Introduction}
Atomic palladium, being the last d element in the 4th row in the periodic chart of the elements, displays a valence that may be a function of the molecule, compound or of the dimensionality of the structure in which it participates. As a free atom it exhibits the electronic configuration of a noble gas but having the 4d shell filled and the 5s5p shells energetically accessible has led several authors to propose that palladium clusters and bulk palladium under pressure could become magnetic. Bulk palladium in its crystalline form and at atmospheric pressure is paramagnetic with a high magnetic susceptibility \cite{blundell}. First principles simulations of palladium under negative pressure \cite{moruzzi} and experimental work in palladium nanoclusters \cite{sampedro} indicate that these samples may display magnetic properties.  Also, the fact that it has a high parameter of Stoner \cite{martin} has made palladium a very appealing subject. Calculations come and go, and results are reported, but experiment should say the final word. Could it be then that amorphous palladium (\textit{a}-Pd) may also display interesting magnetic properties that would shed light on a better understanding of magnetism in bulk materials, both defective and crystalline? The properties of the amorphous phase have been little studied and consequently this is \textit{terra ignota} that must be explored.

Motivated by these considerations, in this work we investigate the effect of topological disorder in the electronic and magnetic properties of amorphous samples of bulk palladium at zero kelvin. We propose that atomic disorder in solid palladium could generate magnetism since this disorder would induce an unbalance in the number of nearest neighbours, locally creating unsaturated bonds leading to a net spin and consequently to a net magnetic moment. This, together with the appearance of holes in the corresponding d band, due to the spilling over of electrons unto the s and p bands may contribute to magnetism. We have performed \textit{ab initio} calculations and the results indicate that, in fact, magnetism may appear in \textit{a}-Pd: Palladium may become an itinerant ferromagnet in the amorphous bulk phase at atmospheric pressure and at T = 0 K. Since our results are obtained for zero temperature, it is reasonable to ask: could it be possible to find this magnetism for non-zero temperature?. We believe the answer is yes, as long as palladium is maintained at very low temperatures in an amorphous state, in a similar manner as superconducting amorphous bismuth exists at T $\lesssim$ 6 K \cite{mata}. 

To computationally generate amorphous structures of palladium we use a technique developed by our group that has given good structures for other materials \cite{valladares2011, mata, valladares2008} where the topology obtained resembles quite accurately the experimental results for the disordered phases of the materials studied; this is the \textit{undermelt-quench} approach. This approach allows the generation of disorder in an otherwise unstable crystalline structure, isodense to the stable one, by heating it to just below the melting temperature of the real material and then cooling it down to the lowest temperature possible. In this manner, a disordered specimen is created and then an optimization run is carried out to release stresses and let the sample reach local equilibrium. Our previous computational studies give us confidence in our procedure and therefore in our present results. However, see Fig. 1 in Ref. \cite{pastukhov2009} where an amorphous phase analogous to the ones we obtained is reported and also compare with the experimental Pd-Pd partial Pair Distribution Function (pPDF) taken from Ref \cite{masumoto, waseda} invoked later on. \\

\begin{figure}[h]
\includegraphics[width=\columnwidth]{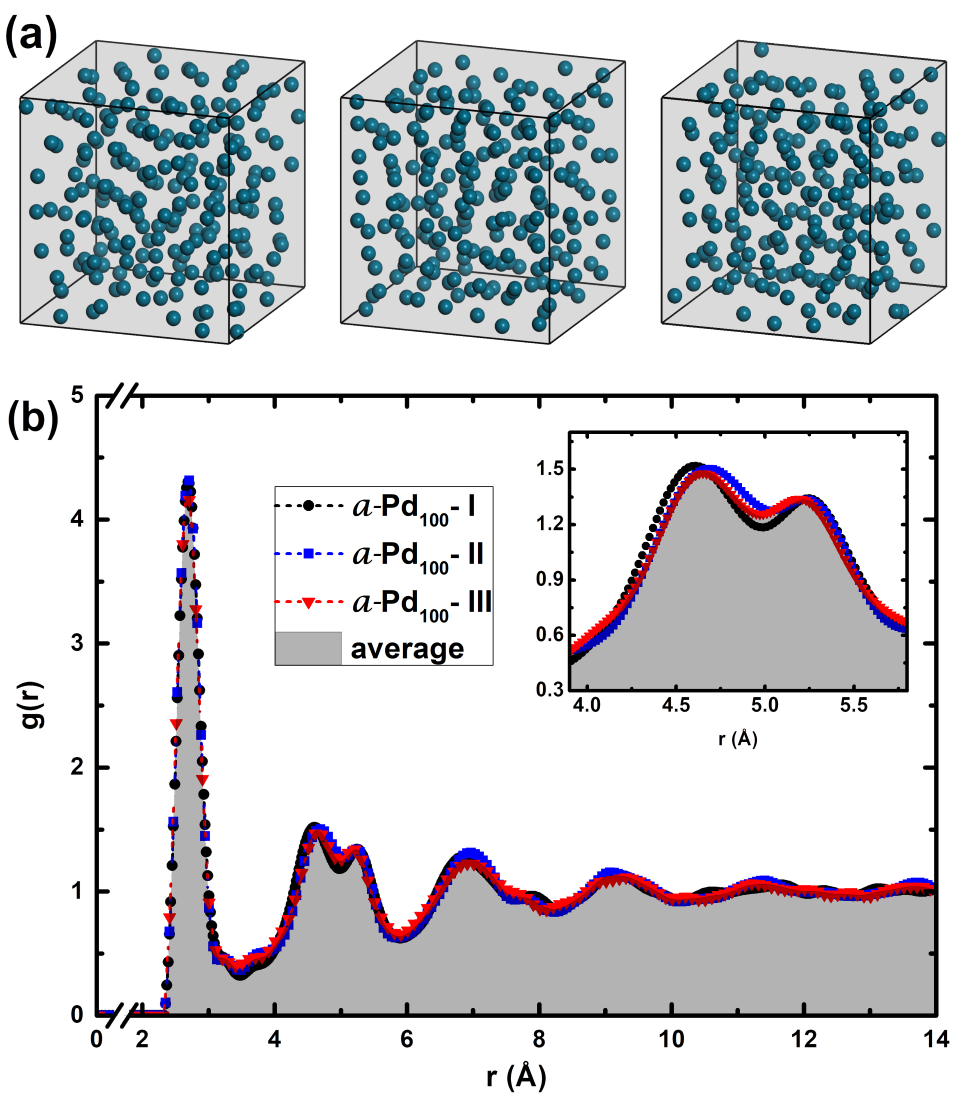}
\caption{\label{fig:fig1} Topology of amorphous palladium; the initially unstable, diamond-like, supercell used contains 216 atoms. (a) Sphere representation of the three atomic structures computationally generated. (b) Pair distribution functions for the three \textit{ab initio} simulated supercells shown in (a). The inset depicts the structure of the bimodal second peak.}
\end{figure}

A variation of this approach consists in doing molecular dynamics also on an unstable specimen at a given constant temperature, under or over the melting point of the real material. The optimization (relaxation) run then ensues. For Pd we did precisely this on a supercell with 216 atoms, and generated, using \textit{ab initio} techniques, three amorphous structures presented in Fig. \ref{fig:fig1}(a) with atoms represented by spheres and whose Pair Distribution Functions (PDFs or $g(r)$) are shown in Fig. \ref{fig:fig1}(b). The bimodal structure of the ‘second’ peak, typical of amorphous metallic elements, can be seen in the inset. See also Ref. \cite{pastukhov2009}.

\section{Method}
The computational tools utilized are contained in the suite of codes Materials Studio (MS) \cite{MS}. In particular, to perform the Molecular Dynamics (MD) and the Geometry Optimization (GO), and to calculate the electronic and magnetic properties of \textit{a}-Pd, the code CASTEP was used \cite{castep}.

\begin{figure*}
\includegraphics[width=0.95\textwidth]{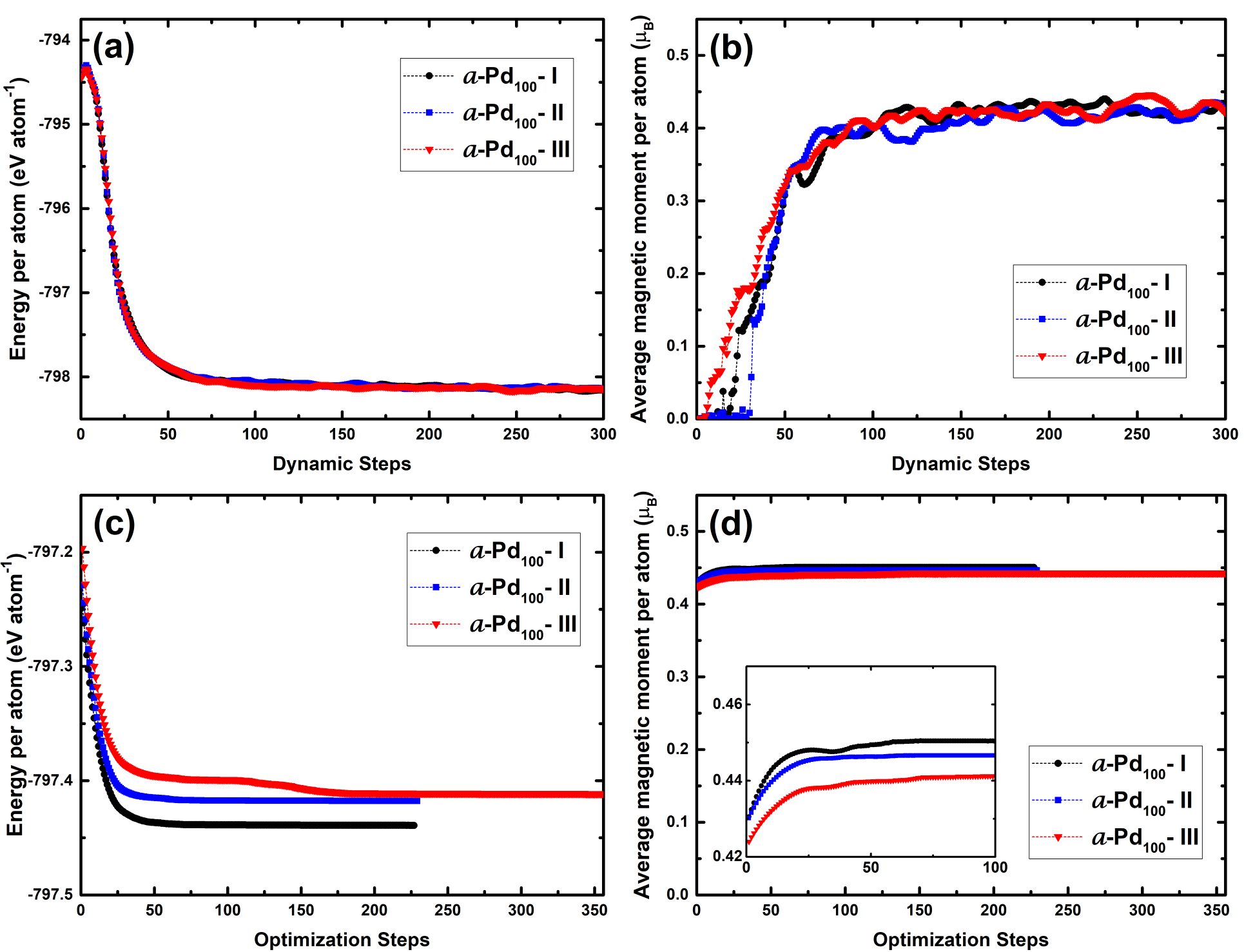}
\caption{\label{fig:fig2}Molecular dynamics and geometry optimizations for the three palladium supercells. (a) Energy per atom as a function of steps of MD. (b) AMM per atom (in Bohr magnetons $\mu_B$) as a function of the MD steps. (c) Energy per atom as a function of the GO steps. (d) AMM per atom (in Bohr magnetons $\mu_B$) per step of GO (tends to 0.45 $\mu_B$ per atom). The inset details the behaviour in the first 100 steps. }
\end{figure*}

A crystalline palladium supercell of 216 atoms was constructed with diamond symmetry (unstable) and with the experimental crystalline density of 12.0 g/cm$^3$; this instability allowed the \textit{undermelt-quench} process to generate amorphous supercells, as mentioned in refs \cite{valladares2011, mata, valladares2008}. The cell underwent three independent MD processes. Once they were complete the three resulting structures were subjected each to a GO procedure starting with a total spin of 93, 96 and 97 $\mu_B$ generated by the MD on each cell. The results indicate that the final topological structures, determined through the PDF, are essentially the same, Fig. \ref{fig:fig1}(b). The energy and the Average Magnetic Moment (AMM), in Bohr magnetons $\mu_B$, are shown in Fig. \ref{fig:fig2}.

For the NVT MD the following approximations were used. The PBEsol functional with a zero spin initially; an electron energy convergence tolerance of 2 x 10$^{-6}$ eV with a convergence window of 3 consecutive steps; a cutoff energy of 260 eV to generate the plane-wave basis to represent the 2160 electrons (10 per atom) distributed in 1297 bands (217 empty); a process at 1,500 K using a thermal bath controlled by a Nose-Hoover thermostat, with a time step of 5 fs during 300 steps, for a total duration of 1.5 ps, and a Pulay mixing scheme. To optimize the MD process, the palladium ultrasoft pseudopotential, Pd\_00PBE.usp, included in the MS suite of codes was the choice, Figs. \ref{fig:fig2}(a) and \ref{fig:fig2}(b).

For the GO the following parameters were employed. The minimization of the energy of the structure was performed with the density mixing method under the Pulay scheme; the functional PBEsol and the relativistic treatment according to Koelling-Harmon included in MS; the energy cutoff for the plane waves was set to 300 eV; the 2160 electrons (10 per atom) were distributed in 1353 bands (226 empty). The initial spin was the output of the MD results: 93, 96 and 97 $\mu_B$; an electron energy convergence tolerance of 1 x 10$^{-6}$ eV with a convergence window of 2 consecutive steps and a smearing of 0.1 eV; the geometry energy tolerance used was 1 x 10$^{-5}$ eV; the force tolerance used was 3 x 10$^{-2}$ eV/Å; the displacement tolerance used was 2 x 10$^{-3}$ \AA\ and the geometry stress tolerance was set to 5 x 10$^{-2}$ GPa. Here the maximum number of steps was set to 1,000 to make sure it would relax within the tolerance limits. The energy systematically diminishes until an arrangement of atoms in local equilibrium is reached, Figs. \ref{fig:fig2}(c) and \ref{fig:fig2}(d).

To investigate the magnetic properties, we ran both the MD and GO processes with unrestricted spin, so the magnetism would evolve freely and acquire a value congruent with a minimum energy structure. The magnetic moment per atom begins to manifest in the first 50 steps of MD and increases until the end of the run, Fig. \ref{fig:fig2}(b). Afterwards, it increases somewhat during the GO process and tends to a constant value, 0.45 $\mu_B$ per atom, Fig. \ref{fig:fig2}(d). The inset shows details of the first 100 steps of GO. 

\begin{figure*}
\includegraphics[width=0.95\textwidth]{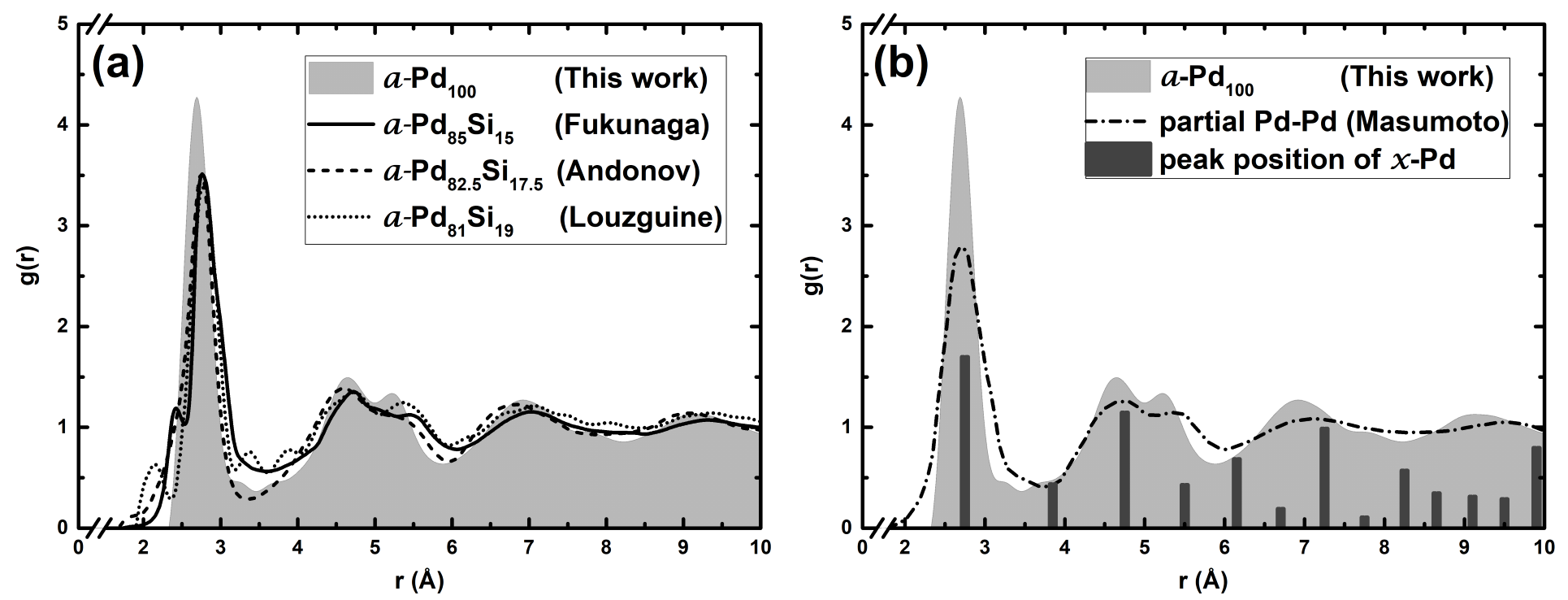}
\caption{\label{fig:fig3}Comparison between total and partial experimental and simulated (solid grey) PDFs. (a) PDF for the simulated \textit{a}-Pd and for the total experimental \textit{a}-PdSi alloys. (b) PDF for the simulated \textit{a}-Pd, partial Pd-Pd obtained by Masumoto \textit{et al}. \cite{masumoto} and PDF for the crystalline structure. The agreement between our simulations and the experiment by Masumoto and coworkers (as reported in ref  \cite{waseda}) is impressive.}
\end{figure*}

How can we be sure that the PDFs obtained do represent the amorphous structure of bulk palladium and that therefore the AMM obtained corresponds to the amorphous phase? We could argue that since our previous results \cite{valladares2011, mata, valladares2008} are very close to the experimental ones, the PDFs that we report in this work should be adequate to describe \textit{a}-Pd; however, since to our knowledge nobody has experimentally produced the pure amorphous phase, we decided to validate our topological findings by using some experiments reported in the literature. Experimentalists have obtained PDFs for amorphous palladium-silicon alloys: Masumoto and coworkers \cite{masumoto} studied these alloys and determined a Pd-Pd pPDF, reported in ref \cite{waseda} and reproduced in Fig. \ref{fig:fig3}(b) in agreement with our simulations. Fukunaga \textit{et al}. \cite{fukunaga} studied \textit{a}-Pd$_{85}$Si$_{15}$, whereas Andonov and collaborators  \cite{andonov} studied \textit{a}-Pd$_{82.5}$Si$_{17.5}$. The system \textit{a}-Pd$_{81}$Si$_{19}$ was studied by Louzguine \cite{louzguine}; all of them are shown in Fig. \ref{fig:fig3}. We first compare the total PDFs they report for the alloys with our simulated PDF for the pure, Fig. \ref{fig:fig3}(a); where the similarities can be observed. We next compare, in Fig. \ref{fig:fig3}(b), our simulation with the experimental result by Masumoto \textit{et al}. \cite{masumoto, waseda} for a Pd-Pd pPDF; the agreement is spectacular. For reference purposes the peaks that describe the atomic positions in crystalline Pd (\textit{x}-Pd) are also presented. The simulated PDF for the pure is the average value displayed in Fig. \ref{fig:fig1}(b). A more detailed study of \textit{a}-PdSi alloys is in the making (I.R. \textit{et al}. Manuscript in preparation). 

\begin{figure*}
\includegraphics[width=0.95\textwidth]{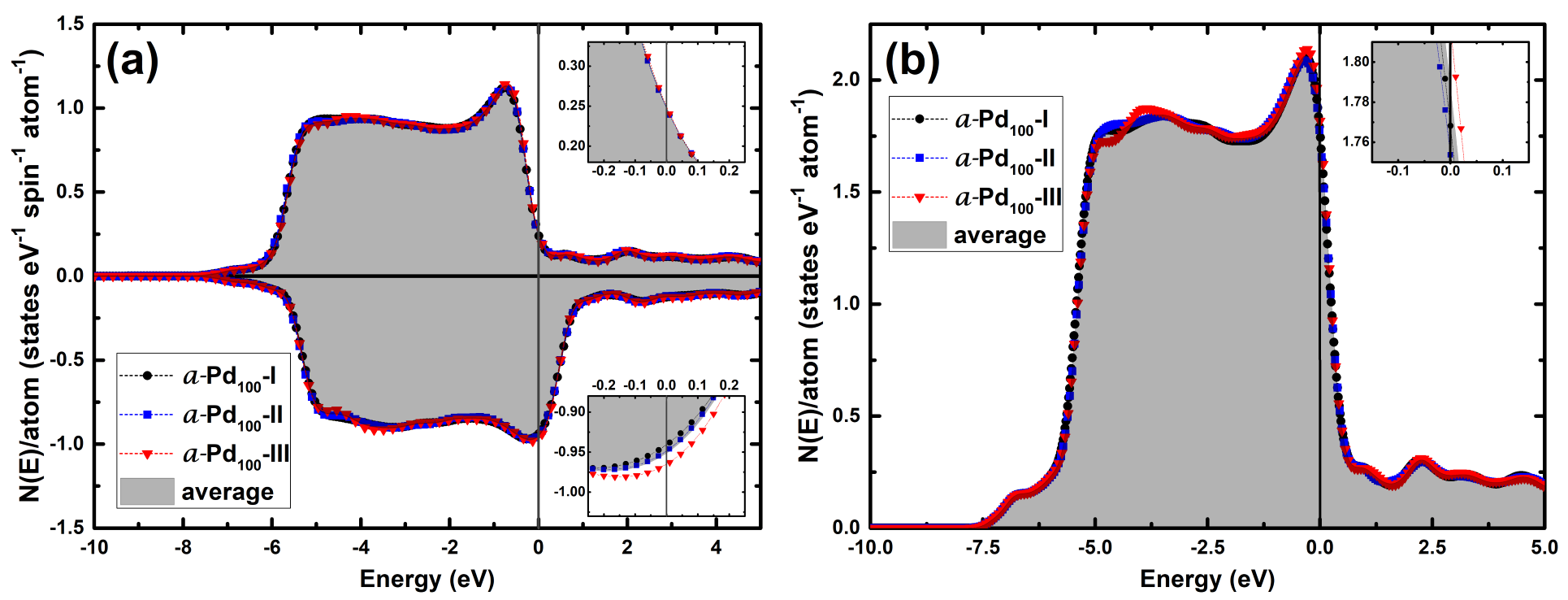}
\caption{\label{fig:fig4}Calculated densities of states for our three ab initio simulated supercells of \textit{a}-Pd. (a) for $\alpha$ and $\beta$ spins; a non-zero magnetism appears when the two types of spins are contrasted, indicating a net magnetic moment. (b) for the non-magnetic state (the number of unpaired electrons is set equal to zero at the start of a single point energy calculation). The average is solid grey. The insets show the details at the Fermi level. }
\end{figure*}

But what about the magnetic properties discovered in our simulations? Is this topological structure indicative of some exciting, non-expected, electronic or magnetic properties of \textit{a}-Pd? If we calculate the number of nearest neighbours (nn) by integrating the area under the first peak of the PDF, we could infer that something is going on since it is smaller than 12, the number of nn in the crystalline fcc phase. This, together with the overflow of electrons from the d shell to the sp shells may be an indicator of an unexpected behaviour. However, since identifying unambiguously the cutoff value to calculate the nn in amorphous metals is a controversial subject \cite{waseda} we opted for a complementary, direct approach in a manner similar to our previous calculations on bismuth \cite{valladares2018, rodriguez}, and obtained the densities of electronic states with $\alpha$ spins and with $\beta$ spins to see if they indicate a net magnetic moment, and they do, Fig. \ref{fig:fig4}(a). For this we also used CASTEP \cite{castep} in the suite of codes of Materials Studio \cite{MS}.

\section{Results and discussion}
To corroborate our results, we did some testing as follows. We calculated the average energy per atom, magnetic and non-magnetic, and we found that the non-magnetic value was -797.418 eV/atom and for the magnetic structure was -797.438 eV/atom. The magnetic structure is more stable and the difference of 0.02 eV/atom is of the order of reported results for some silicon phases, 0.016 eV/atom in going from silicon diamond, the stable phase, to hexagonal diamond \cite{yin}. We also carried out some \textit{ab initio} computational calculations for the crystalline unit cells of nickel (fcc) and iron (bcc), both with zero spin and with non-zero spin initially, using the same code (CASTEP) and the same parameters as for the palladium jobs, to test our results and procedures. When the initial spin was zero, our code and our approach led to non-magnetic results for both materials; this suggested that a magnetic trigger may be needed. When values of spin of 1 and 2 $\mu_B$ per atom were assigned to nickel and 1 and 4 $\mu_B$ to iron, the energy optimization run gave a net magnetic moment of 0.68 $\mu_B$ per atom of nickel, for both runs, and 2.47 $\mu_B$ per atom of iron, for both runs. Compare these results to experiment: 0.61 $\mu_B$ per atom for Ni \cite{mook} and 2.22 $\mu_B$ per atom for Fe \cite{shull}. We performed similar runs for a unit cell of gold and found no magnetism with or without an initial magnetic trigger. We also ran amorphous 216-atom supercells of copper, silver and gold, with and without an initial magnetic trigger, and found no remnant magnetism. This indicates that if our procedure is applied to all these materials it leads to the expected behaviour. We conclude that all these results validate our findings for amorphous palladium.

\begin{table*}
\caption{\label{tab:tab1}Energy, magnetism and Stoner criterion($UN(E_F)$) for the three amorphous palladium supercells studied.}
\begin{ruledtabular}
\begin{tabular}{lcccccc}
\multirow{2}{*}{System}&\multicolumn{2}{c}{Total energy per atom $\left (\frac{\text{eV}}{\text{atom}} \right )$}&\multirow{2}{*}{AMM ($\mu_B$)}&\multirow{2}{*}{$\Delta E$ $\left (\frac{\text{eV}}{\text{atom}} \right )$}&\multirow{2}{*}{$\delta E$ $\left (\frac{\text{eV}}{\text{atom}} \right )$}&\multirow{2}{*}{$UN(E_F)$} \\[2pt] \cline{2-3} \\[-8pt]
&Magnetic&Non-magnetic&&&& \\[2pt] \hline \\[-6pt]
\textit{a}-Pd$_{100}$-I&-797.4391&-797.4178&0.45&-0.021&0.22&1.50\\[2pt]
\textit{a}-Pd$_{100}$-II&-797.4323&-797.4178&0.45&-0.015&0.23&1.31\\[2pt]
\textit{a}-Pd$_{100}$-III&-797.4427&-797.4189&0.44&-0.024&0.23&1.50\\[2pt]
Average&-797.4380&-797.4181&0.45&-0.020&0.23&1.44
\end{tabular}
\end{ruledtabular}
\end{table*}

Our electronic calculations indicate that the overflow of electrons, invoked for crystalline Pd, exists also for the amorphous 5s, 5p and 4d states: 4d$^{10-x}$ (5s5p)$^{x}$; however, we claim that the spin band splitting in the absence of a magnetic field will be more preponderant in amorphous Pd than in crystalline Pd and that the energy balance $\Delta E = K (1 - UN(E_F))$ [with $K = (1/2) N(E_F)\delta E^2$ and $U = \mu_0{\mu_B}^2\lambda$ (see Ref. \cite{blundell} p. 145)]  will now be $\Delta E = K (1 - UN(E_F) - Vf_B)$ where $V$ indicates the contribution to the magnetic splitting of the unsaturated bonds $f_B$  in the amorphous. This heuristic argument would lead us to a modified Stoner criterion for the stability of the \textit{a}-Pd magnetic phase: $[UN(E_F) + Vf_B] \geq 1$, and the spontaneous ferromagnetism is possible for smaller values of $UN(E_F)$.

To quantify the traditional Stoner criterion, $UN(E_F) \geq 1$, for the spontaneous spin band splitting we need to obtain the product $UN(E_F)$ from our computational results. Here we must mention that in what follows, our numerical values are a consequence of the parameters and approximations used in our simulations; in particular, the use of the PBEsol functional, and as such the values reported herein may differ from others where different functionals are used. The variations of energetic and geometrical results depend on the functionals used in atoms and molecules \cite{pople}, although no magnetism is considered in this reference. This comment also applies to the results reported in Table 1.

First, we start with the equation for the total energy change between the magnetic and non-magnetic states $\Delta E$, fifth column in Table 1 (see ref  \cite{blundell} p. 146):

\begin{equation}
\Delta E = \frac{1}{2} N(E_F)(\delta E)^2 \left [1 -UN(E_F)\right ],
\end{equation}

\noindent where $N(E_F)$ is the non-magnetic result (obtained by setting the number of unpaired electrons equal to zero at the outset of an energy calculation)  and $\delta E$ is the difference between the highest energies for the magnetic and non-magnetic free electron gas. The average energy difference is $\Delta E =$ –0.02 eV atom$^{-1}$ as shown in Table \ref{tab:tab1}.

The product $UN(E_F)$ then becomes:
\begin{equation}
UN(E_F)= 1- \left [ \frac{2\Delta E}{N(E_F) (\delta E)^2} \right ].
\end{equation}

To calculate $\delta E$ we first obtain the value of the proportionality constant $\gamma$ in the expression for the density of states for the free electron gas in three dimensions $N(E)= \gamma \sqrt{E}$ by requiring that the integral from the bottom of the band to $E_F$ ($E_F=$ 7.87 eV  for the non-magnetic state in Figure \ref{fig:fig4}(b)) integrates to 10 states eV$^{-1}$ atom$^{-1}$.

\begin{equation}
\int_{0}^{E_F} N(E_F) dE = \int_{0}^{E_F} \gamma \sqrt{E}dE = 10;
\end{equation}

\noindent therefore the proportionality constant becomes $\gamma$ = 0.68 eV$^{-3/2}$. Next we evaluate the areas under the spin-up and spin-down curves in Figure \ref{fig:fig4}(a), 4.78 states per atom for beta and 5.22 states per atom for alpha,  and then map them onto the free electron parabola to obtain the unbalance at the Fermi energy, $\delta E$.

Once we have these results and the total density of states at the Fermi level for the non-magnetic state (1.78 states eV$^{-1}$ spin$^{-1}$ atom$^{-1}$ in Figure \ref{fig:fig4}(b)) we then obtain an average value of $UN(E_F)$ = 1.44 which, being larger than 1, satisfies the Stoner criterion (Table \ref{tab:tab1}).

\section{Conclusions}
The results reported indicate that the magnetic state is more stable than the non-magnetic. Also, we generated an amorphous structure for palladium that agrees with available experimental, partial, results. By looking at the densities of electronic states we conclude that amorphous Pd continues being a metal; in fact, a metallic glass. The Stoner criterion holds and therefore we surmise that the amorphous phase is an itinerant ferromagnet. So that the validity of our results can be assessed, the 1.44 value obtained for $UN(E_F)$ should be compared to those found for iron: 1.43 \cite{janak}, or nickel: 2.03 \cite{janak}, or even crystalline palladium: 0.78 \cite{janak}. These findings may open a novel field in the magnetism of defective metals such that, when macro defects are considered like pores or voids, it may be useful in industry to produce light weight strong magnets.
Evidently, no calculation can force a material to behave in a certain manner, so the final judge is the experiment.  Recent experimental advances, commented in Ref  \cite{greer}, discuss the possibility of obtaining pure amorphous metals and these efforts may well be the beginning of a whole new field. Other comments that should be kept in mind when dealing with simulations are those of Ref \cite{zunger} to avoid some of the pitfalls discussed there. We believe we have been extremely careful not to force the simulations to produce a specific outcome. Finally, since our results are obtained for zero temperature, this magnetism may exists at low temperatures as long as Pd remains amorphous.

\begin{acknowledgments}
I.R. and D.H.R. acknowledge CONACyT for supporting their graduate studies. A.A.V., R.M.V. and A.V. thank DGAPA-UNAM for continued financial support to carry out research projects under grant IN104617. M.T. V\'azquez and O. Jim\'enez provided the information requested. Simulations were partially carried out in the Computing Center of DGTIC-UNAM.
\end{acknowledgments}

\bibliographystyle{apsrev4-1}
\bibliography{PRB_bib.bib}

\end{document}